# Diffusion and drift of graphene flake on graphite surface

Irina V. Lebedeva[1,2,3,1)], Andrey A. Knizhnik[2,3,2)], Andrey M. Popov[4,3)], Olga V. Ershova[1,5], Yurii E. Lozovik[4,1,4)], and Boris V. Potapkin[2,3]

[1]Moscow Institute of Physics and Technology, 141701, Institutskii pereulok 9, Dolgoprudny, Moscow Region, Russia

[2]RRC "Kurchatov Institute", 123182, Kurchatov Sq. 1, Moscow, Russia

[3]Kintech Lab Ltd, 123182, Kurchatov Sq. 1, Moscow, Russia

[4]Institute of Spectroscopy, 142190, Fizicheskaia Str. 5, Troitsk, Russia

[5]School of Chemistry, University of Nottingham, University Park, Nottingham NG7 2RD, United Kingdom

**ABSTRACT**

Diffusion and drift of a graphene flake on a graphite surface are analyzed. A potential energy relief of the graphene flake is computed using *ab initio* and empirical calculations. Based on the analysis of this relief, different mechanisms of diffusion and drift of the graphene flake on the graphite surface are considered. A new mechanism of diffusion and drift of the flake is proposed. According

---

[1)] Electronic mail: lebedeva@kintechlab.com.

[2)] Electronic mail: knizhnik@kintechlab.com.

[3)] Electronic mail: am-popov@isan.troitsk.ru.

[4)] Electronic mail: lozovik@isan.troitsk.ru.

to the proposed mechanism, rotational transition of the flake from commensurate to incommensurate state takes place with subsequent simultaneous rotation and translational motion until a commensurate state is reached again, and so on. Analytic expressions for the diffusion coefficient and mobility of the flake corresponding to different mechanisms are derived in wide ranges of temperatures and sizes of the flake. The molecular dynamics simulations and estimates based on *ab initio* and empirical calculations demonstrate that the proposed mechanism can be dominant under certain conditions. The influence of structural defects on the diffusion of the flake is examined on the basis of calculations of the potential energy relief and molecular dynamics simulations. The methods of control over the diffusion and drift of graphene components in nanoelectromechanical systems are discussed. The possibility to experimentally determine the barriers to relative motion of graphene layers based on the study of diffusion of a graphene flake is considered. The results obtained can be also applied to polycyclic aromatic molecules on graphene and should be qualitatively valid for a set of commensurate adsorbate-adsorbent systems.

**I. INTRODUCTION**

Due to the unique electronic and mechanical properties[1] of graphene, it is considered as a promising material for a variety of applications. Intensive studies of relative rotational and translational quasistatic motion of graphene layers are currently carried out[2-12]. Particularly, one of the most interesting phenomenon for graphene goes by the name of "superlubricity", i. e., the ultra-low static friction between incommensurate graphene layers[3-12]. Here we for the first time study dynamic behavior of a graphene flake on a graphite surface. Namely, we show that anomalous fast diffusion and drift of a graphene flake on a graphene layer or a graphite surface are possible through rotation of the flake to incommensurate states. As opposed to superlubricity observed in *non-equilibrium* systems (such as a flake moved by the tip of the friction force microscope), diffusion and drift refer to the behavior of the system *close to thermodynamic equilibrium*.

The relative motion of flat nanoobjects is determined by the potential energy relief, i.e. the dependence of the interaction energy between the nanoobject and surface on three coordinates, two of which correspond to the position of the center of mass and the third one is the rotation angle with respect to the surface. The quasistatic superlubricity is observed when the motion takes place across the nearly flat potential energy relief. The fact that this phenomenon is observed for graphene is related to features of the potential energy relief for a graphene flake on a graphite surface. At particular rotation angles, the lattice vectors of the flake can be chosen similar to those of the underlying graphene layer. In these states, the flake is commensurate with the graphite surface and the potential energy relief is highly non-uniform, with significant barriers to the motion of the flake. At other rotation angles, the flake is incommensurate with the graphite surface and the energy of the flake almost does not depend on its position, i.e. there are no barriers to its motion. Such incommensurate states are observed in the form of so-called Moiré patterns[13,14]. It is seen that the incommensurate states of the flake should be "superlubric". In the present work, we suggest that transition of the flake to the incommensurate states affects not only the *static* tribological behavior of the graphene flake *attached to the tip of the friction force microscope* but also the *dynamic* behavior of the free flake and thus might provide anomalous fast diffusion and high mobility of the flake.

The mechanism of ultra-low static friction related to the structural incompatibility of contacting surfaces was first suggested by Hirano and Shinjo[15,16]. Later, Dienwiebel *et al.*[3-5] measured the friction between a graphene flake attached to the tip of the friction force microscope and a graphite surface. For most of the orientations of the graphene flake, Dienwiebel *et al.* observed very low friction forces, while for some orientations corresponding to the commensurate states, the tip performed a so-called stick-slip motion and the measured friction forces were significantly higher.

Theoretical studies[5-12] were performed to investigate the dependences of the static friction between graphene layers on the pulling direction, stacking structure, interlayer distance, loading force and structural defects. The density functional theory (DFT) was applied so far only to calculate the variation of the interlayer interaction energy and force in graphite upon relative sliding of the layers[12,17]. In the present work, we perform DFT calculations of the dependence of the interlayer interaction energy not only on the relative position but also on the relative orientation of graphene layers.

Incommensurability in adsorbate-adsorbent systems is known to result in fast diffusion of the adsorbate. Experimentally it was demonstrated that large non-epitaxially oriented gold or antimony clusters can diffuse on a graphite surface with a surprising diffusion coefficient of about $10^{-8}$ cm$^2$/s at room temperature[18,19], which is significantly higher than the diffusion coefficients for clusters epitaxially oriented on the surface (of the order of $10^{-17}$ cm$^2$/s[20,21]). Molecular dynamics (MD) simulations[22-24] confirmed that the fast diffusivity of a cluster on a surface originates from its interfacial incommensurability. However, such a fast diffusivity was found up to now only for systems where the adsorbate and adsorbent are *not commensurate* at the ground state. Based on the systematic study of diffusion mechanisms for a graphene flake on a graphite surface, we suggest that a diffusion mechanism through rotation of the adsorbate to the "superlubric" incommensurate states (see FIG. 1) is possible in adsorbate-adsorbent systems which are interfacially *commensurate* at the ground state. *Ab initio* and empirical calculations performed here show that the barrier for rotation of the graphene flake to incommensurate states has the same order of magnitude as the barrier for transitions between adjacent energy minima in the commensurate states. Therefore, there should be a competition between the diffusion mechanisms through transitions of the flake between adjacent energy minima in the commensurate states and through rotation of the flake to the incommensurate states. The estimates and simulations demonstrate that under certain conditions, the proposed diffusion mechanism through rotation of the flake to the incommensurate states is dominant. We believe that a similar mechanism of diffusion should also be prominent in any other commensurate adsorbate-adsorbent systems. Particularly, the results obtained here can be useful for understanding of dynamics of polycyclic aromatic molecules on graphene.

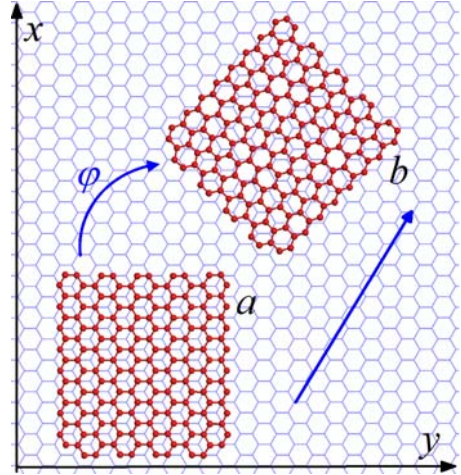

FIG. 1. Schematic representation of the proposed diffusion mechanism through rotation of the flake to the incommensurate states. Structure of the graphene flake on the graphene layer: (a) commensurate state, (b) incommensurate state.

The diffusion and drift properties of real graphene flakes can be influenced by the presence of atomic scale defects in the flakes. Recent advances in microscopic technologies have made it possible to see a wide range of defects in graphene and carbon nanotubes at the atomic scale[25–27]. Among the most studied defects, there are Stone–Wales[28] (5-7-5-7 defect, with four adjacent hexagons switching to two pentagons and two heptagons) and vacancy[29] defects. The 5-7 defects play an important role in the formation of carbon nanotube junctions[30,31] and for plastic deformations of carbon nanotubes[32-34] and graphene[35]. The Stone–Wales and vacancy defects were demonstrated to modify the static friction between graphene layers[8]. The defects were also shown to increase the dissipation and the level of fluctuations in the nanotube-based gigahertz oscillators[36-38]. However, the influence of defects on the dynamic behavior of graphene flakes has not been studied yet. Here we examine the effect of defects in the flake both on the static potential energy relief of the graphene flake and its diffusion characteristics.

In addition to the fundamental problems discussed above, the study of diffusion and drift characteristics of graphene is also of interest with regard to the use of graphene flakes in nanoelectromechanical systems (NEMS)[39]. Ability of free relative motion of graphene layers[2] allows using them as movable elements of NEMS similar to walls of carbon nanotubes[36,40-45]. Because of the small size of NEMS, such systems are subject to significant thermodynamic fluctuations[36,40,45]. On the one hand, relative diffusion[46] or only random displacement[41] of NEMS components due to thermodynamic fluctuations can disturb the NEMS operation[41,46]. On the other hand, the diffusion can be used in Brownian motors[44]. Another crucial issue in nanotechnology is the possibility to control the motion of NEMS components with an external force[42,43]. We suggest that fixation of the orientation of a graphene flake can facilitate or hinder diffusion and drift of the flake and propose to use such a fixation in graphene-based NEMS. Possible ways to fix the orientation of a graphene flake are discussed.

The study of diffusion of a graphene flake on a graphite surface can also help to determine the barrier for relative motion of graphene layers, which is a problem of high importance for elaboration of graphene-based and nanotube-based NEMS. Up to now, only upper limits of shear strength were measured for relative motion of carbon nanotube walls[47], while only the order of magnitude was estimated for relative motion of graphene layers[2]. As for theoretical studies, the available data on the barriers to relative motion of walls of the (5,5)@(10,10) nanotube demonstrate that the results of calculations based on different *ab initio* and empirical methods differ by several orders of magnitude (see Table 1 of Ref. [48] and references therein). We propose that experimental measurements of the diffusion coefficient of a graphene flake on a graphite surface can provide the *true* value of the barrier for relative motion of graphene layers.

The paper is organized in the following way. In Sec. II, we describe the model which is used in our calculations and investigate the dependences of the interaction energy for a graphene flake on a graphite surface on the position and orientation of the flake. The potential energy reliefs of the flake obtained using empirical and *ab initio* methods are compared. Based on these studies, the possible mechanisms of diffusion and drift of the graphene flake on the graphite surface are discussed. Sec. III is devoted to MD simulations demonstrating the proposed diffusion mechanism of the graphene flake. The diffusion coefficients for the free flake and for the flake with the fixed commensurate orientation are obtained and compared at different temperatures. In Sec. IV and Sec. V, we derive analytic expressions for the diffusion coefficient and mobility of the free flake and of the flake with the fixed commensurate orientation. The contributions of different drift and diffusion mechanisms of the flake are analyzed for different temperatures and sizes of the flake. In Sec. VI, we propose the possible methods of control over the diffusion and drift of NEMS components using electric and magnetic fields which affect the orientation of these components. The required field strengths needed for the effective control are estimated. In Sec. VII, we study the potential energy relief of the flake with structural defects and investigate the influence of the defects on the diffusion characteristics of the flake by MD simulations. Our conclusions are summarized in Sec. VIII.

## II. ANALYSIS OF POTENTIAL ENERGY RELIEF

To analyze the possible mechanisms of diffusion and drift of a graphene flake on a graphite surface we calculate the potential energy relief for this system. We also compare the results obtained using empirical and *ab initio* methods. On the basis of this study, we discuss possible mechanisms of diffusion and drift of the flake.

As a model system for energy calculations and MD simulations, we consider a rectangular graphene flake placed on an infinite graphene layer (see FIG. 1). Such a bilayer system is appropriate for studying diffusion of a flake on multilayer graphene or graphite as well, since the interaction of non-adjacent graphene layers is much weaker than the interaction of the adjacent ones[49]. To model the infinite substrate layer the periodic boundary conditions are applied along mutually perpendicular armchair and zigzag directions. The size of the graphene flake is 2.0 nm along the armchair edge and 2.1 nm along the zigzag edge (178 carbon atoms). The size of the model cell is 5.5 nm x 5.7 nm, respectively. The interaction between atoms of the graphene flake and the underlying graphene layer at distance $r$ is described by the Lennard–Jones 12–6 potential $U_{LJ}(r) = 4\varepsilon\left((\sigma/r)^{12} - (\sigma/r)^{6}\right)$ with parameters $\varepsilon = 3.73$ meV, $\sigma = 3.40$ Å taken from the AMBER database[50] for aromatic carbon. The cut-off distance of the Lennard–Jones potential is 20 Å. The Lennard-Jones potential provides the interlayer binding energy in graphite of –62 meV/atom, which is consistent with the experimental value of –52±5 meV/atom obtained from the recent experiments on thermal desorption of polycyclic aromatic hydrocarbons from a graphite surface[51]. The covalent carbon-carbon interactions in the layers are described by the empirical Brenner potential[52], which has been shown to correctly reproduce the vibrational spectra of carbon nanotubes[53] and graphene nanoribbons[54] and has been widely applied to study carbon systems[36,40,43,55].

To calculate the dependences of the interlayer interaction energy on the position and orientation of the graphene flake, the structures of the flake and graphene layer are separately relaxed using the Brenner potential and then the flake is rigidly shifted and rotated parallel to the underlying graphene layer. The distance between the flake and the infinite graphene layer is 3.4 Å. Account of structure deformation induced by the interlayer interaction was shown to be inessential for the shape of the potential relief for the interaction between graphene-like layers, such as the interwall interaction of carbon nanotubes[56,57] and the intershell interaction of carbon nanoparticles[58,59]. For example, the account of the structure deformation of the shells of the $C_{60}@C_{240}$ nanoparticle gives rise to changes of the barriers for relative rotation of the shells which are less than 1% [58,59].

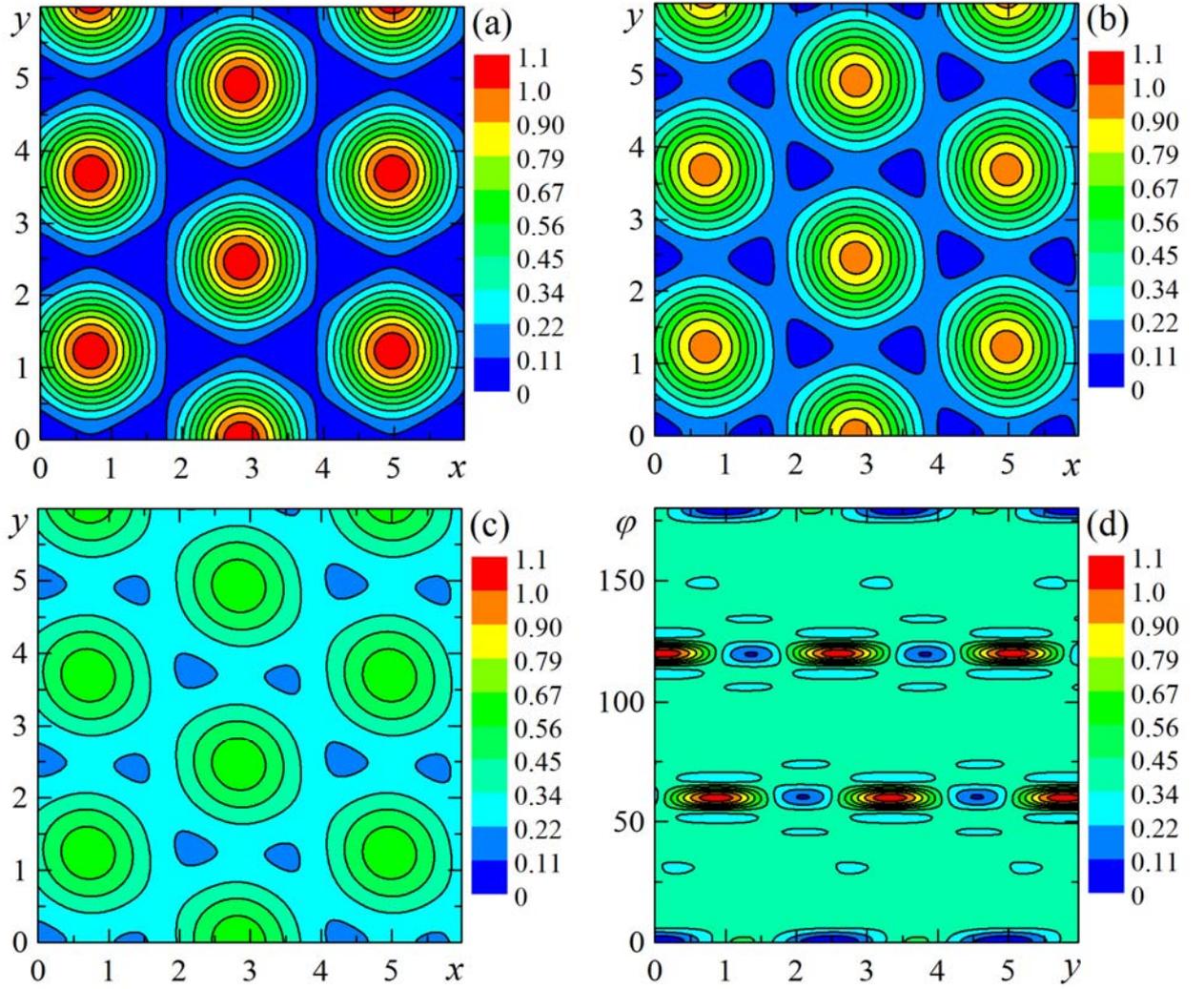

FIG. 2. The interlayer interaction energy between the graphene flake and the graphene layer (in meV/atom) calculated using the Lennard–Jones potential as a function of the position of the center of mass of the flake $x, y$ (in Å, $x$ and $y$ axes are chosen along the armchair and zigzag directions, respectively) and its rotation angle $\varphi$ (in degrees). (a) $\varphi = 0°$, (b) $\varphi = 2°$, (c) $\varphi = 4°$, (d) $x = 0$. The energy is given relative to the energy minimum.

The calculated interaction energy between the flake and the surface as a function of its position and orientation is shown in FIG. 2. The rotation angle $\varphi$ is measured relative to the commensurate orientation of the flake, so that $\varphi = 0°, 60°, 120°$, etc. are attributed to the commensurate states. All other rotation angles correspond to the incommensurate states. The found minimum energy states of the flake correspond to the commensurate AB-stacking (see FIG. 1a), in agreement with Ref. [60]. There is the energy barrier $\varepsilon_{com} = 0.10$ meV/atom for transition of the flake between adjacent energy minima. However, even at temperatures above this energy barrier, a long-distance free motion of the flake is not possible due to the numerous energy hills on the potential energy relief, which are higher than the energy barrier $\varepsilon_{com}$ by an order of magnitude. The maximum energy states of the flake correspond to the AA-stacking. The energy difference between the AA and AB-stackings is found to be $\varepsilon_{max} = 1.1$ meV/atom.

Based on the approximation[10,61] for the interaction of a single carbon atom in the graphene flake with the graphite surface containing only the first Fourier components, it is easy to show that the potential energy relief for the flake in the commensurate states can be roughly approximated in the form

$$U = U_1 \left( cos(2k_1 x) - 2 cos(k_1 x) cos(k_2 y) \right) + U_0, \qquad (1)$$

where $k_2 = 2\pi / a_0$ and $k_1 = k_2 / \sqrt{3}$, $x$ and $y$ axes are chosen along the armchair and zigzag directions, respectively (see FIG. 1). The parameters $U_1 = 0.225$ meV/atom and $U_0 = -61.92$ meV/atom are fitted to reproduce the potential energy relief calculated using the Lennard–Jones potential. For these parameters, the root-mean square deviation of the potential energy relief (1) from that calculated using the Lennard–Jones potential equals $0.15 U_1$.

It is seen from FIG. 2 (see the full version of FIG. 2 in S1, Ref. 62) that with rotation of the graphene flake, the magnitude of corrugation of the potential energy relief decreases. At the angle of $10°$, the magnitude of this corrugation is negligibly small (less than $0.25 U_1$). At the angle of $60°$, the flake becomes again commensurate with the graphene layer. We find the width of the energy wells and energy peaks in the dependence of the interlayer interaction energy on the orientation of the flake to be about $2\delta\varphi \approx 2a_0/L$ (see FIG. 2d and FIG. 3), where $a_0 = 2.46$ Å is the lattice constant for graphene and $L$ is the size of the flake, in agreement with Refs. [3–5, 10]. The energy of the incommensurate states relative to the commensurate ones (which is equal to the energy needed to rotate the flake by the angle $\delta\varphi \approx a_0/L$) is seen to be $\varepsilon_{in} = 0.37$ meV/atom (see FIG. 2d and FIG. 3).

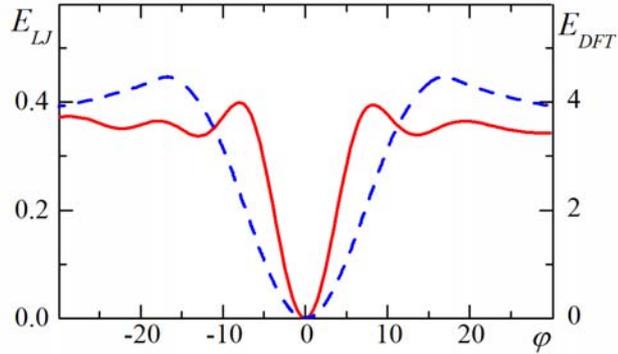

FIG. 3. The interlayer interaction energy between the graphene flake and the graphene layer (in meV/atom) calculated using the Lennard–Jones potential for the 178-atom flake (red solid line; left axis) and DFT for the 54-atom flake (blue dashed line; right axis) as a function of the rotation angle $\varphi$ (in degrees). The value $\varphi = 0°$ corresponds to the global energy minimum.

Since the Lennard-Jones potential was claimed not to be sensitive enough to the relative position of the graphene layers[17], we perform similar calculations on the basis of DFT. The VASP code[63] with the local density approximation (LDA)[64] is used. The basis set consists of plane waves with the maximum kinetic energy of 358 eV. The interaction of valence electrons with atomic cores is described using ultrasoft pseudopotentials[65]. The error in calculations of the system energy is less than 0.005 meV/atom for the chosen cutoff energy of the plane waves. Integration over the Brillouin zone is performed using a single k-point sampling. For the DFT calculations, we consider a smaller system. The graphene flake consists of 54 carbon atoms and has all edges terminated with hydrogen atoms. The size of the model cell is 2.0 nm x 2.1 nm x 1.3 nm. The DFT calculations show that for the graphene flake, there is an energy minimum of about −34 meV/atom at the distance of 3.32 Å from the graphene layer. This binding energy is consistent with the experimental value of 35±10 meV/atom for the interlayer binding energy in graphite obtained from the analysis of collapsed multi-walled carbon nanotubes[66] but is smaller that the recent experimental value[51] mentioned above. The DFT calculations of the potential energy relief of the graphene flake are performed for the equilibrium distance of 3.32 Å between the flake and the graphene layer. We do not include a dispersion correction in our DFT calculations. However, studies performed for polycyclic aromatic molecules on a graphene flake showed that the dispersion correction provides a small contribution (<10%) to the energy difference between the AA and AB-stackings [67].

According to the DFT calculations, the minimum energy states of the system also correspond to the commensurate AB-stacking. The energy difference between the AA and AB-stackings and the barrier for motion of the flake from one energy minimum to another are calculated to be $\varepsilon_{max} = 10.77$ meV/atom and $\varepsilon_{com} = 1.28$ meV/atom, respectively, in agreement with the previous DFT calculations[17]. The relative energy of the incommensurate states is found to be $\varepsilon_{in} = 4.0$ meV/atom (see FIG. 3). The potential energy relief obtained by the DFT calculations is approximated using expression (1) with the parameters $U_1 = 2.38$ meV/atom and $U_0 = -30.48$ meV/atom. The root-mean square deviation of the potential energy relief (1) from the calculated relief equals $0.04 U_1$.

Thus, we have found that the shapes of the potential energy reliefs obtained both using the empirical and DFT calculations are described by expression (1) which contains the only the first Fourier components. These shapes are qualitatively the same but the magnitudes of corrugation of the interlayer energy differ by an order of magnitude. Note also that both the empirical [57, 68] and DFT [40, 48] calculations showed that the first Fourier components are sufficient for approximation of potential energy reliefs of carbon nanotubes with commensurate walls. Therefore, the potential relief of the interaction energy of a graphene flake and a graphene layer or a graphite surface can be characterized with a single energy parameter, e.g., $\varepsilon_{com}$ ($\varepsilon_{in} \approx 3.5\varepsilon_{com}$ and $\varepsilon_{max} \approx 10\varepsilon_{com}$), which, however, takes different values for different calculation methods. Note that the energy parameter $\varepsilon_{com}$ has not been yet measured experimentally. Nevertheless, we show below that the diffusion and drift of the flake are mostly determined by the ratio of the energy parameter of the potential energy relief multiplied by the size of the flake to temperature $\varepsilon_{com}N/k_BT$ ($k_B$ is Boltzmann's constant, $T$ is temperature and $N$ is the number of atoms in the flake) rather than by the energy parameter alone. So the results obtained below for certain temperature $T$ and number $N$ of atoms in the flake should be also valid for the systems with the same ratio $\varepsilon_{com}N/k_BT$.

Based on the calculations of the potential energy relief for the graphene flake on the infinite graphene layer (see FIG. 2), we propose that different diffusion and drift mechanisms should be realized depending on temperature and size of the flake. The diffusion of the flake commensurate with the underlying graphene layer at low temperatures $T \ll T_{com}$, where $T_{com} = N\varepsilon_{com}/k_B$, proceeds by rare jumps between adjacent energy minima. On increasing temperature to the region $T_{com} \leq T \ll T_{max}$, where $T_{max} = N\varepsilon_{max}/k_B$, the barriers for transitions of the flake between adjacent energy minima become less than the thermal kinetic energy of the flake. So these barriers are not noticeable for the flake during its motion. However, there are still many high potential energy hills, which serve as scattering centers to motion of the flake in the commensurate states and restrict its diffusion length.

Another diffusion mechanism of the flake should be related to rotation of the flake to the incommensurate states. At temperatures $T \ll T_{in}$, where $T_{in} = N\varepsilon_{in}/k_B$, the probability for the flake to acquire the energy required for rotation to the incommensurate states is small compared to that for transitions between adjacent energy minima in the commensurate states. However, we show below that this factor can be compensated by long distances passed by the flake before it returns to the commensurate states. Furthermore, on increasing temperature, the time spent by the flake in the incommensurate states also increases. Therefore, we predict that there can be a competition between the diffusion mechanisms for the commensurate and incommensurate states of the flake. At temperatures $T \geq T_{\mathrm{in}}$, rotation of the flake becomes almost free, which should provide the dominant contribution of the proposed diffusion mechanism to the diffusion of the flake.

At high temperatures $T \gg T_{\mathrm{max}}$, the magnitude of corrugation of the potential energy relief of the graphene flake becomes small compared to the thermal kinetic energy ($k_{\mathrm{B}}T \gg N\varepsilon_{\mathrm{max}}$). At these temperatures, the difference between the diffusion of the flake in the commensurate and incommensurate states disappears and the diffusion coefficient of the flake should reach its maximum value. The drift mechanisms of the flake at different temperatures should be similar to the diffusion mechanisms.

### III. MOLECULAR DYNAMICS DEMONSTRATION OF FAST DIFFUSION

To demonstrate that diffusion of a graphene flake can actually proceed through rotation of the flake to the incommensurate states, we perform MD simulations of diffusion of the graphene flake on the graphene layer. On the basis of the MD simulations, we obtain and compare the diffusion coefficients of the free flake and of the flake with the fixed commensurate orientation at different temperatures.

The microcanonical MD simulations of diffusion of the graphene flake have been performed at temperatures $T = 50 - 500$ K using the Lennard-Jones potential and the Brenner potential. The flake consisting of 178 atoms is considered. An in-house MD-kMC code has been implemented. The time step is 0.4 fs. The initial configuration of the system is optimized at zero temperature. During the simulations, the substrate layer is fixed at three atoms, while the flake is left unconstrained.

Transitions of the flake through the boundaries of the model cell are properly taken into account in the calculations of the diffusion length.

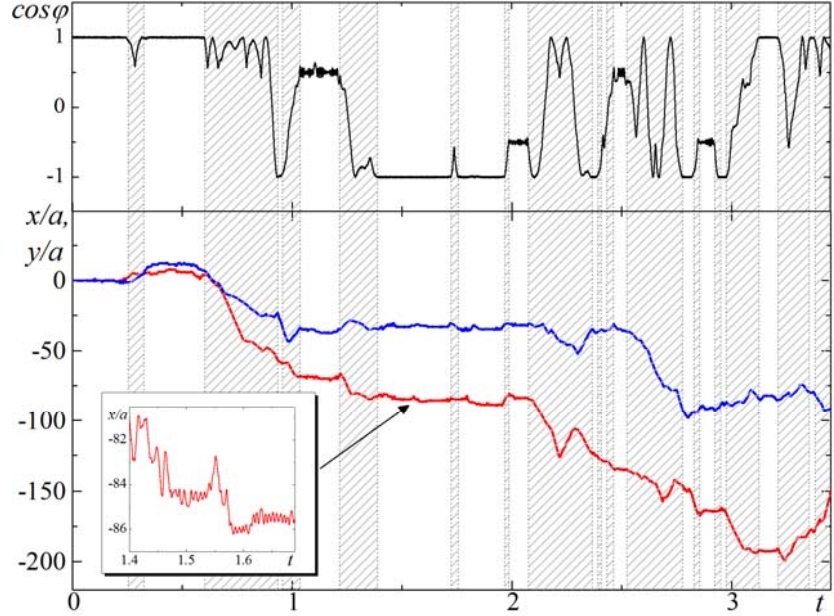

FIG. 4. Calculated position $x/a$ and $y/a$ (red and blue lines, respectively) of the center of mass of the graphene flake and orientation of the flake $\cos\varphi$ (black line) as functions of time $t$ (in ns) at temperature 300 K. The shaded areas correspond to the fast diffusion of the graphene flake in the incommensurate states. Insert: Scaled-up time dependence of coordinate $x/a$ of the center of mass of the graphene flake corresponding to the slow diffusion of the flake in the commensurate states.

Let us first discuss the results obtained at room temperature. For the flake consisting of 178 atoms, temperature 300 K corresponds to $T \approx 1.4 T_{com}$ and lies in the range $T_{com} < T < T_{in}$. Therefore, the competition between diffusion of the flake in the commensurate states with scattering at the potential energy hills and through rotation of the flake to the incommensurate states should be observed at this temperature. The MD simulations of diffusion of the free flake clearly demonstrate that the flake can be found in two states (see FIG. 4). In the first state, the flake stays almost commensurate with the underlying graphene layer and the rotation angle of the flake only slightly oscillates. The transitions of the flake between the energy minima are observed (see the insert of FIG. 4). Nevertheless, the distances passed by the flake across the surface are small and comparable to the lattice constant of graphene. In the second state, as the flake acquires the energy equal to the relative energy of the incommensurate states, it starts the free motion across the surface accompanied by simultaneous rotation.

The diffusion coefficient of the flake at temperature 300 K has been obtained on the basis of 6 simulations of 3-3.5 ns duration. The asymptotic behavior $\langle r(t)^2 \rangle = 4Dt$, where $D$ is the diffusion coefficient, is almost reached at the times considered (see S2, Ref. 62). We estimate the diffusion coefficient at temperature 300 K to be $D = (3.6 \pm 0.5) \cdot 10^{-4}$ cm$^2$/s.

To investigate the behavior of the diffusion coefficient with temperature we have performed the MD simulations of the diffusion of the flake in the temperature range of 50 – 500 K. The diffusion coefficients at temperatures 200 and 500 K calculated on the basis of 4 simulations of 1-2 ns duration are given in TABLE I. The same as at 300 K, both diffusion mechanisms through transitions of the flake between adjacent energy minima in the commensurate states and through rotation of the flake to the incommensurate states contribute to the diffusion of the flake at these temperatures. At temperatures 50 and 100 K, no rotation to the incommensurate states is detected within the simulation time of a few nanoseconds. Our estimates show that the time required to observe such a rotation at temperatures below 100 K exceeds $0.1\,\mu$s, which is beyond the reach of our MD simulations due to the limited computer resources. Nevertheless, the fact that these events are rare and cannot be observed in the MD simulations does not necessarily mean that they do not provide a noticeable contribution to the diffusion of the flake. So the diffusion coefficients calculated at temperatures 50 and 100 K should be attributed *only to the diffusion in the commensurate states* and are listed in TABLE II. The contribution of the diffusion mechanism through rotation of the flake in the incommensurate states at these temperatures is estimated in Sec. IV.

TABLE I. Calculated diffusion coefficient $D$, average time $\langle \tau_{rot} \rangle$ of rotation by the angle $\Delta\varphi \approx \pi/3$, average time $\langle \tau_{st} \rangle$ of stay in the commensurate states between these rotations and mean-square distance $\langle l^2 \rangle$ passed by the flake as it rotates by the angle $\Delta\varphi$ for the free graphene flake at different temperatures.

| $T$ (K) | $D$ (cm$^2$/s) | $\langle \tau_{st} \rangle$ (ps) | $\langle \tau_{rot} \rangle$ (ps) | $\langle l^2 \rangle$ (Å$^2$) |
|---|---|---|---|---|
| 200 | $(2.9 \pm 1.7) \cdot 10^{-4}$ | $11 \pm 3$ | $42 \pm 12$ | $300 \pm 100$ |
| 300 | $(3.6 \pm 0.5) \cdot 10^{-4}$ | $13.4 \pm 1.2$ | $20.0 \pm 0.9$ | $240 \pm 30$ |
| 500 | $(7 \pm 3) \cdot 10^{-4}$ | $9.0 \pm 1.1$ | $18.0 \pm 1.5$ | $280 \pm 60$ |

TABLE II. Calculated diffusion coefficient $D_c$ and average time $\langle \tau \rangle$ between transitions of the flake from one energy minimum to another for the graphene flake with the fixed commensurate orientation at different temperatures.

| $T$ (K) | $D_c$ (cm$^2$/s) | $\langle \tau \rangle$ (ps) |
|---|---|---|
| 50 | $(2.6 \pm 0.5) \cdot 10^{-7}$ | $196 \pm 34$ |
| 100 | $(1.09 \pm 0.11) \cdot 10^{-6}$ | $47 \pm 5$ |
| 200 | $(5.3 \pm 0.5) \cdot 10^{-6}$ | $9.5 \pm 0.8$ |
| 300 | $(6.7 \pm 0.6) \cdot 10^{-6}$ | $9.0 \pm 1.0$ |

To clarify the relative contributions of different diffusion mechanisms to the diffusion of the flake, we have performed the MD simulations of the diffusion of the flake with the fixed commensurate orientation. The diffusion coefficients $D_c$ calculated on the basis of two to three simulations of 1-2 ns duration at temperatures 50 – 300 K are given in TABLE II. From comparison of TABLE I and TABLE II, it follows that at temperatures 200 – 500 K, the diffusion coefficient $D_c$ corresponding to the diffusion of the flake only in the commensurate states is orders of magnitude smaller than the total diffusion coefficient $D$. This proves that the proposed diffusion mechanism through rotation of the flake in the incommensurate states should provide the most significant contribution to the diffusion of the flake under these conditions.

To get a deeper insight into the diffusion mechanisms, we analyze the trajectories of the flake (see FIG. 4). In the simulations of the diffusion of the free flake, it is assumed that the flake reaches the commensurate state as soon as the rotation angle of the flake gets in the interval $\varphi_0 - \delta\varphi < \varphi < \varphi_0 + \delta\varphi$ ($\varphi_0 = 0°, 60°$, etc.). The average time $\langle \tau_{rot} \rangle$ of rotation by the angle $\Delta\varphi \approx \pi/3 \gg 2\delta\varphi$, the average time $\langle \tau_{st} \rangle$ of stay in the commensurate states between these rotations and the mean-square distance $\langle l^2 \rangle$ passed by the flake as it rotates by the angle $\Delta\varphi$ calculated at different temperatures are given in TABLE I. It is seen from TABLE I that in the temperature range of 200 – 500 K, the flake spends most of the time in the incommensurate states ($\langle \tau_{rot} \rangle / \langle \tau_{st} \rangle = 2 - 4$). This is related to abundance of the incommensurate states compared to the commensurate ones. The average distance $l$ passed by the flake during its rotation by the angle $\Delta\varphi$ considerably exceeds the distance $a = a_0/\sqrt{3} = 1.42$ Å between adjacent energy minima in the commensurate states (see TABLE I). Further analysis shows that when the flake passes the commensurate state, it can be trapped in a potential energy well (see FIG. 3), reflect from a potential energy hill (see FIG. 2) or continue its rotation in the same direction. Assuming that the flake is trapped in the commensurate state if it stays there for time longer than $\tau_0' = 6.2$ ps (which corresponds to the period of small rotational vibration in the potential well shown in FIG. 3), the probability for the flake to get trapped in the commensurate states is found to be about 0.27. The probabilities for the flake to reflect from the potential energy hills (see FIG. 2) and to pass the commensurate states are estimated to be about 0.17 and 0.56, respectively. Though the probability for the flake to get trapped in the commensurate states is rather low, the linear velocity of the flake is noticeably changed (by the value $|\Delta \vec{V}| \sim V$, where $V$ is the linear velocity of the flake) almost every time the flake passes the commensurate states, restricting the diffusion length of the flake. For simplicity, we assume that the translational motion of the flake is strongly disturbed every time as it rotates by the angle $\Delta \varphi \approx \pi / 3$ in our estimates in Sec. IV. The simulations for the flake with the fixed commensurate orientation show that the diffusion of such a flake proceeds by transitions of the flake between adjacent energy minima at the distance $a = a_0 / \sqrt{3} = 1.42$ Å from each other. The average time $\langle \tau \rangle$ between these transitions is given in TABLE II.

## IV. ESTIMATES OF DIFFUSION COEFFICIENT

Let us derive analytic expressions for the diffusion coefficients of the free flake and of the flake with the fixed commensurate orientation and analyze contributions of different diffusion mechanisms to the total diffusion coefficient of the free flake for different temperatures and sizes of the flake. We use the following expression for the diffusion coefficient of the flake in the two-dimensional case

$$D = \frac{\langle l^2 \rangle}{4 \langle \tau \rangle}, \tag{2}$$

where $\langle l^2 \rangle$ is the mean-square distance passed by the flake in a single diffusion step and $\langle \tau \rangle$ is the average time corresponding to a single diffusion step.

We start our consideration from diffusion of the flake with the fixed commensurate orientation. At temperatures $T \ll T_{\max}$, the diffusion of the flake in the commensurate states is possible only by transitions between adjacent energy minima. In this case, the diffusion length of the flake in the commensurate state equals the distance $l = a_0/\sqrt{3}$ between adjacent energy minima. The rate constant $k(T)$ for transitions of the flake between adjacent energy minima is calculated numerically in the framework of the transition state theory (see S3 Ref. 62, Ref. 69 and references therein). We find that it can be interpolated as $k(T) \approx 2.1\tau_0^{-1} exp(-N\varepsilon_{com}/k_BT)$ at temperatures $T < 0.25T_{com}$ and $k(T) \approx 0.39\tau_0^{-1}(k_BT/N\varepsilon_{com})^{0.916}$ at temperatures $T > 1.5T_{com}$, where $\tau_0 \approx 5.8\,\text{ps}$ is the period of small translational vibrations of the flake about the energy minimum. Note that following the discussion of Sec. II, the function depends on the ratio of the energy parameter characterizing the potential energy relief of the flake multiplied by the size of the flake to temperature $\varepsilon_{com}N/k_BT$. Based on the expressions for the diffusion length and for the average time corresponding to a single diffusion step $\langle \tau \rangle = 1/k$, we find the diffusion coefficient of the flake with the fixed commensurate orientation according to formula (2)

$$D_c = \frac{a_0^2 k(T)}{12}. \tag{3}$$

At room temperature for the flake consisting of 178 atoms, this diffusion coefficient equals $D_c \approx 9 \cdot 10^{-6}\,\text{cm}^2/\text{s}$, which is close to the diffusion coefficient estimated on the basis of the MD simulations for the flake in the commensurate state (see TABLE II).

The behavior of the flake in the incommensurate states is mostly determined by dynamic friction and random thermal forces. Therefore, to analyze the diffusion of the free flake we first study the characteristics of dynamic friction between the graphene flake and graphite surface using microcanonical MD simulations in the temperature range of 50 – 300 K. In these simulations, the flake is given an initial linear or angular velocity so that the kinetic energy of the flake is about 10 eV, which is much greater than the thermal kinetic energy and the magnitude of corrugation of the potential energy relief for the flake. The linear and angular velocity correlation times $\tau_c$ and $\tau_c'$ (which are the characteristic decay times of the linear and angular velocity autocorrelation functions $\langle V_\alpha(t)V_\beta(t+\Delta t)\rangle_t = \delta_{\alpha\beta}\langle V_\alpha^2(t)\rangle_t exp(-\Delta t/\tau_c)$ and $\langle \omega(t)\omega(t+\Delta t)\rangle_t = \langle \omega^2(t)\rangle_t exp(-\Delta t/\tau_c')$, where $\alpha$ and $\beta$ denote the projections on axes $x$ and $y$) are calculated on the basis of 100 simulations of 10 ps duration for each set of conditions (temperature, rotation angle of the flake and direction of the linear velocity for the case of translational motion). The calculated dependences of the correlation times on temperature, the direction of the linear velocity and the orientation of the flake are shown in FIG. 5. It is seen that $\tau_c \approx \tau_c'$ and the linear velocity correlation time weakly depends on the direction of the linear velocity and the orientation of the flake. In our further calculations, we use the linear approximation for the temperature dependence of the reciprocals of the velocity correlation times presented in FIG. 5

$$\tau_c^{-1} \approx \tau_c'^{-1} \approx \left(7.5\cdot 10^{-3}T[K]+1.72\right)\cdot 10^9 \text{s}^{-1}. \qquad (4)$$

At room temperature, the linear and angular velocity correlation times are found to be approximately $\tau_c \approx \tau_c' \approx 250$ ps.

Let us now analyze the diffusion mechanisms of the free flake. While the flake stays in the commensurate state, the translational and rotational motions of the flake are determined by three forces: potential force, dynamic friction force and random force. As the flake leaves the commensurate state, the potential force becomes negligibly small compared to the friction and random forces and can be omitted. The rotational and translational motions of the flake while it rotates by the angle $\Delta\varphi \approx \pi/3$ can proceed either in ballistic or diffusive regimes depending on the relation between the time of rotation by this angle $\tau_{rot}$ and the linear and angular velocity correlation times $\tau_c' \approx \tau_c$. In the ballistic regime ($\tau_{rot} \ll \tau_c' \approx \tau_c$), the friction between the flake and underlying graphene layer and the thermal noise can be disregarded, i.e. the linear and angular velocities of the flake can be assumed virtually constant. In the diffusive regime ($\tau_{rot} \gg \tau_c' \approx \tau_c$), the rotational and translational motions of the flake during a single diffusion step are damped and should be described using diffusion equations with the diffusion coefficients $D_\varphi = k_B T \tau_c' / I$ and $D_t = k_B T \tau_c / M$ (where $M = Nm$ is the mass of the flake, $I \propto N^2 m a_0^2$ is the moment of inertia of the flake), respectively, following from the Einstein relation. As soon as the flake reaches again a commensurate state, the potential force becomes relevant. Due to the increased dissipation of the rotational energy and the energy exchange between the rotational and translational degrees of freedom in the commensurate state, the flake can be trapped. However, even if the flake continues its rotation, the potential force disturbs the translational motion of the flake, which limits the diffusion length of the flake in the ballistic regime. In the diffusive regime, the diffusion length of the flake is limited by friction.

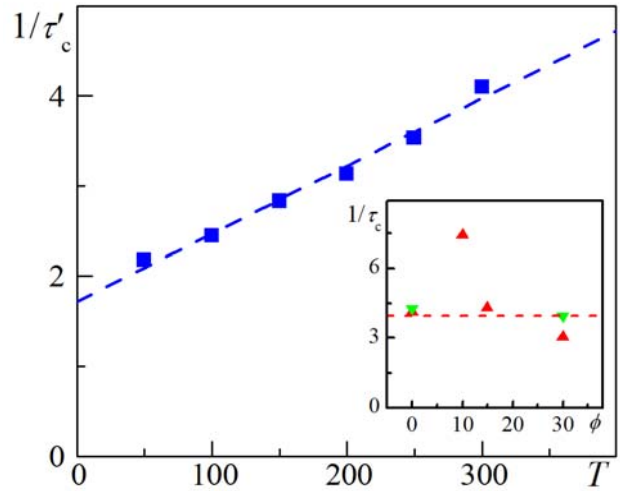

FIG. 5. Calculated reciprocal of the angular velocity correlation time $\tau_c'$ (in ns) as a function of temperature $T$ (in K). The dashed line shows the linear approximation of the obtained data. Insert: Calculated reciprocal of the linear velocity correlation time $\tau_c$ (in ns) as a function of the angle $\phi$ (in degrees) of the linear velocity of the flake with respect to the armchair direction (axis $x$) for the commensurate (▲) and incommensurate ($\varphi = 15^0$, ▼) states of the flake at temperature 300 K. The dashed line corresponds to the reciprocal of the angular velocity correlation time.

As shown by the MD simulations (see Sec. III), for the flake consisting of 178 atoms, the ballistic regime is realized in the wide temperature range. To estimate the contribution of the diffusion mechanism through rotation of the flake to the incommensurate states to the total diffusion coefficient of the free flake in the case when the rotation proceeds in the ballistic regime ($\tau_{rot} \ll \tau'_c \approx \tau_c$), we assume that the translational motion of the flake is strongly disturbed every time the flake passes the commensurate states and a single diffusion step corresponds to rotation by the angle $\Delta\varphi$, which is supported by the MD simulations. Furthermore, in the ballistic regime ($\tau_{rot} \ll \tau'_c \approx \tau_c$), the linear and angular velocities of the flake do not change significantly within the time of rotation from one commensurate state to another. On the basis of the equilibrium Maxwell–Boltzmann distributions for the linear and angular velocities of the flake, $V$ and $\omega$, we find (see S4, Ref. 62) the average time of rotation by the angle $\Delta\varphi \approx \pi/3$ and the mean-square distance passed by the flake while it rotates by the angle $\Delta\varphi$ to be

$$\langle \tau_{rot} \rangle \approx \frac{\sqrt{\pi}\Delta\varphi}{\omega_T}, \qquad \langle l^2 \rangle \approx \frac{2V_T^2 \Delta\varphi^2}{\omega_T^2} ln\left(\frac{\omega_T \tau'_c}{\Delta\varphi}\right) \qquad (5)$$

where $\omega_T = \sqrt{2k_B T/I}$ is the thermal angular velocity of the flake and $V_T = \sqrt{2k_B T/M}$ is the thermal linear velocity of the flake. At room temperature for the considered flake, $\omega_T \approx 5.5 \cdot 10^{10}$ s$^{-1}$, $V_T \approx 4.8 \cdot 10^3$ cm/s and $\langle \tau_{rot} \rangle \approx 22$ ps, in agreement with the result of the MD simulations (see TABLE I). Note that $\langle \tau_{rot} \rangle \ll \tau'_c$, i.e. the condition of the ballistic rotation is satisfied for the flake under consideration at room temperature (see Eq. (4)). The mean-square distance passed by the flake while it rotates by the angle $\Delta\varphi$ at room temperature equals $l \approx 10a = 1.5$ nm, in agreement with the result of the MD simulations (see TABLE I). This quantity considerably exceeds the distance between adjacent energy minima $a = a_0/\sqrt{3}$, which corresponds to the diffusion length for the flake in the commensurate states.

The total average time corresponding to a single diffusion step for the diffusion of the flake through rotation to the incommensurate states is given by the sum of $\langle \tau_{rot} \rangle$, which characterizes the time spent by the flake in the incommensurate states in a single diffusion step, and of the average time of stay in the commensurate states $\langle \tau_{st} \rangle$ between the rotations by the angle $\Delta\varphi$. The average time of stay in the commensurate states can be found from thermodynamic considerations (see S4, Ref. 62) $\langle \tau_{st} \rangle / \langle \tau_{rot} \rangle = \phi(T)$, where function $\phi(T)$ is calculated numerically. At temperatures $T < 0.28 T_{in} \approx T_{com}$, the function $\phi(T)$ exponentially decreases with temperature $\phi(T) \approx 0.17 (k_B T / N\varepsilon_{in})^{1.67} exp(N\varepsilon_{in} / k_B T)$. At temperatures $T > 0.8 T_{in} \approx 2.6 T_{com}$, the temperature dependence of the function $\phi(T)$ is weak and the function reaches about $\phi(T) \approx 0.25$. Note that following the discussion of Sec. II, the function $\phi(T)$ depends on the ratio of the energy parameter characterizing the potential energy relief of the flake multiplied by the size of the flake to temperature $\varepsilon_{in} N / k_B T \approx 3.5 \varepsilon_{com} N / k_B T$. From the above expressions for the function $\phi(T)$ and Eq. (5), we find that at room temperature $\langle \tau_{st} \rangle \approx 11\,\text{ps}$, in agreement with the result of the MD simulations (see TABLE I).

Based on the values of the diffusion length and average time corresponding to a single diffusion step $\langle \tau \rangle = \langle \tau_{st} \rangle + \langle \tau_{rot} \rangle$, we estimate the contribution of the proposed diffusion mechanism through rotation of the flake to the incommensurate states to the total diffusion coefficient of the free flake using expression (2)

$$D_{ib} \approx \frac{\Delta\varphi}{2\sqrt{\pi}} \frac{V_T^2}{\omega_T (1 + \phi(T))} ln\left(\frac{\omega_T \tau_c'}{\Delta\varphi}\right). \tag{6}$$

At room temperature, the formula gives $D_{ib} \approx 1.7 \cdot 10^{-4}\,\text{cm}^2/\text{s}$, in reasonable agreement with the result of the MD simulations (see TABLE I). The difference by the factor of ~ 2 from the value obtained on the basis of the MD simulations is related to some probability for the flake to skip the commensurate state without disturbing the translational motion of the flake.

The contribution of diffusion of the flake in the commensurate states to the total diffusion coefficient of the free flake can be found on the basis of equation (3) for the diffusion coefficient of the flake with the fixed commensurate orientation. However, it should be taken into account that the free flake stays in the commensurate states only for the fraction of time $\alpha_c = \langle \tau_{st} \rangle / (\langle \tau_{rot} \rangle + \langle \tau_{st} \rangle) = \phi / (1+\phi)$. Therefore, the actual contribution of the diffusion of the flake in the commensurate states to the total diffusion coefficient is given by relation $D_{c1} = \alpha_c D_c$. The fraction $\alpha_c$ is close to unity at low temperatures ($T \ll T_{com}$) and decreases with increasing temperature. This means that the contribution of the diffusion mechanism for the flake in the commensurate states to the total diffusion coefficient is even smaller than $D_c$. At room temperature for the flake consisting about 200 atoms, the contribution of this diffusion mechanism is only $0.3 D_c$. The total diffusion coefficient resulting from the diffusion both in the commensurate and incommensurate states can be found as $D = D_{ib} + D_{c1} = D_{ib} + \phi D_c / (1+\phi)$.

The ratio of the contributions of the diffusion mechanisms $D_{ib} / D_{c1}$ and the dependences of the diffusion coefficients $D$ and $D_c$ for the free flake and for the flake with the fixed commensurate orientation on temperature $T / T_{com} = k_B T / N \varepsilon_{com}$ obtained using expressions (3) and (6) are shown in FIG. 6. Let us use FIG. 6 to discuss the diffusion mechanisms of the flake at different temperatures. At low temperatures $T \ll T_{com}$, the contributions of different diffusion mechanisms $D_{ib}$ and $D_{c1}$ exponentially depend on the reciprocal of temperature (which is provided by functions $k(T)$ and $\phi(T)$ in expressions (3) and (6)). Since the barrier $N\varepsilon_{com}$ for transitions of the flake between adjacent energy minima in the commensurate state is much smaller than the energy $N\varepsilon_{in}$ required for rotation of the flake to the incommensurate states $N\varepsilon_{\text{com}} \ll N\varepsilon_{\text{in}}$, at these temperatures the flake can stay only in the commensurate state and jump between adjacent energy minima. At $T \sim T_{\text{com}}$, the temperature dependences of $D_{\text{ib}}$ and $D_{\text{c1}}$ switch from the exponential ones to the dependences weaker than linear ones (see Eqs. (3) and (6)). The diffusion mechanism through rotation to the incommensurate states becomes dominant and the ratio of the contributions of the diffusion mechanisms $D_{\text{ib}} / D_{\text{c1}}$ reaches 10-100 (see FIG. 6a). This is provided both by the decrease of the time spent in the commensurate states and the long distances passed by the flake in the incommensurate states. As a result, in the temperature range of $T \sim (1 \div 3)T_{\text{com}}$, the diffusion coefficient $D$ of the free flake is greater by one-two orders of magnitude than the diffusion coefficient $D_{\text{c}}$ of the flake with the fixed commensurate orientation (see FIG. 6b). However, the translational motion of the flake is still disturbed as it passes the commensurate states. So the diffusion coefficient is still lowered compared to the maximum diffusion coefficient $D_{\text{t}} = k_{\text{B}} T \tau_{\text{c}} / M$ determined by friction. Only at temperatures $T \sim T_{\text{max}} \approx 10 T_{\text{com}}$, the diffusion coefficient of the flake reaches its ultimate value $D_{\text{t}}$. It is also seen from FIG. 6b that the diffusion coefficients $D_{\text{ib}}$ and $D_{\text{c}}$ estimated on the basis of Eqs. (3) and (6) are in agreement with the results of the MD simulations at different temperatures (see TABLE I and TABLE II).

As shown above, the magnitudes of corrugation of the potential relief of the interlayer interaction energy between the graphene flake and the graphene layer obtained through the DFT calculations and using the empirical potentials differ by an order of magnitude. However, even using the data obtained by the DFT calculations, the results of our MD simulations and analytic estimates can still be assigned to flakes of smaller size or at higher temperature. From FIG. 6 and Eqs. (3) and (6), it is seen that the diffusion coefficients $D_{\text{ib}}$ and $D_{\text{c1}}$ depend on the energy parameters of the interlayer interaction in graphite via the factor $T / T_{\text{com}} = k_{\text{B}} T / N \varepsilon_{\text{com}}$, as discussed above. Therefore, relying on the results of the DFT calculations, it can be, for example, shown that diffusion of a flake consisting of 70 atoms at room temperature should proceeds mostly through its rotation to the incommensurate states ($D_{ib} / D_{c1} \sim 10$).

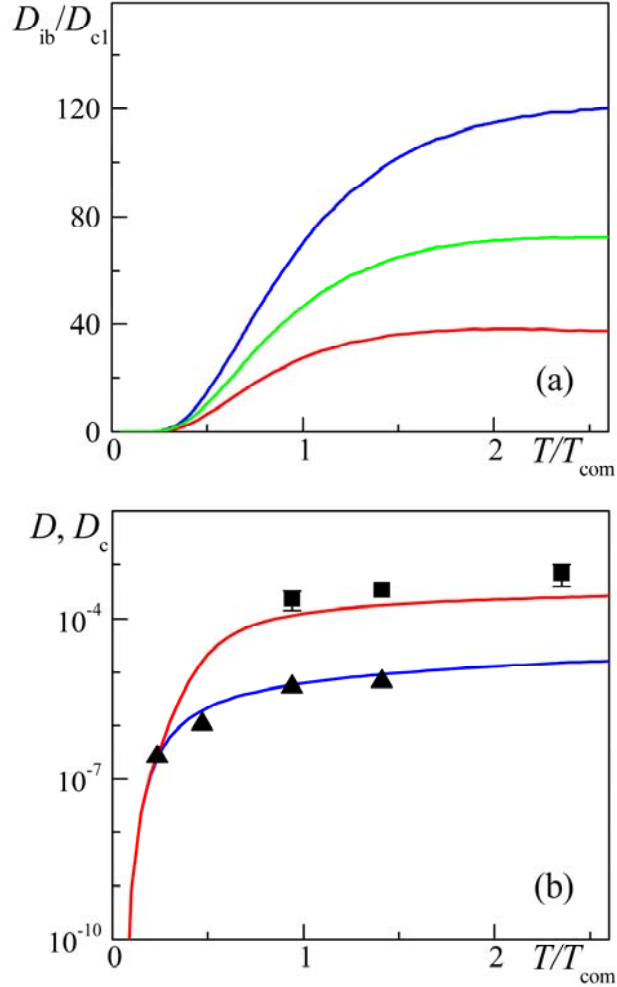

FIG. 6. (a) Calculated ratio $D_{ib} / D_{c1}$ of the contributions of the diffusion mechanisms though rotation of the flake to the incommensurate states, $D_{ib}$, and through transitions of the flake between adjacent energy minima in the commensurate states, $D_{c1}$, to the total diffusion coefficient $D$ of the free flake as a function of temperature $T / T_{com}$ for different sizes of the flake: $N = 40$ (upper blue line), $N = 178$ (middle green line), $N = 700$ (lower red line). (b) Calculated total diffusion coefficient $D$ (red line; in cm$^2$/s) of the free flake and diffusion coefficient $D_c$ (blue line; in cm$^2$/s)

of the flake with the fixed commensurate orientation as functions of temperature $T/T_{\text{com}}$ for $N = 178$. The results of the MD simulations are shown with black squares for the total diffusion coefficient $D$ of the free flake and with black triangles for the diffusion coefficient $D_c$ of the flake with the fixed commensurate orientation.

The case of diffusive rotation of the flake ($\tau_{\text{rot}} \gg \tau_c'$) can be adequate for description of the flake diffusion at high or very low temperatures (according to Eqs. (4) and (5), $\tau_c' \propto (aT+b)^{-1}$ and $\tau_{\text{rot}} \propto T^{-1/2}$, respectively) or for large sizes of the flake ($\tau_{\text{rot}} \propto N$ and $\tau_c'$ does not depend on $N$). In this case, while the flake stays in the incommensurate states, its motion is diffusive with the diffusion coefficient $D_t = k_B T \tau_c / M = V_T^2 \tau_c / 2$ and the mean-square distance passed by the flake is simply given by $\langle l^2 \rangle = 4 D_t \langle \tau_{\text{rot}} \rangle$. Taking into account that $\langle \tau_{\text{st}} \rangle / \langle \tau_{\text{rot}} \rangle = \phi(T)$, we find the diffusion coefficient corresponding to the proposed mechanism of diffusion through rotation to the incommensurate states in the case of diffusive rotation of the flake to be $D_{\text{id}} \approx 0.5 V_T^2 \tau_c / (1 + \phi(T))$.

## V. ESTIMATES OF MOBILITY

We now derive analytic expressions for the mobilities of the free flake and of the flake with the fixed commensurate orientation. The external force leads to a slope of the potential energy relief of the flake, so that the motion in the direction of the force becomes preferable over the motion in other directions. The mobility of the flake can be found by general formula

$$\mu = \frac{\langle l_F \rangle}{\langle \tau \rangle F}, \tag{7}$$

where $F$ is the small external force, $\langle l_F \rangle$ is the average displacement of the flake and $\langle \tau \rangle$ is the average time corresponding to a single drift step. The disturbance of the system behavior under the action of the external force is small as long as conditions $Fa_0 \ll k_B T$ and $F\tau_c \ll \sqrt{2k_B T M}$ are satisfied.

The drift mechanisms of the flake exposed to an external force are similar to the diffusion mechanisms of the flake. Let us first consider drift of the flake with the fixed commensurate orientation. At temperatures $T \ll T_{max}$, the drift is provided by transitions of the flake between adjacent energy minima. Under the action of the force, the barrier for motion of the flake increases by the work of the force $Fa_0 \cos\theta / 2\sqrt{3}$, where $\theta$ is the angle between the directions of the transition and the force. As a result, the average distance passed by the flake in the direction of the force in a single drift step is given by

$$\langle l_F \rangle \approx \frac{a_0}{\sqrt{3}}\left(\langle \cos\theta_i \rangle + \frac{Fa_0}{2\sqrt{3}k_B T}\langle \cos^2\theta_i \rangle\right) = \frac{Fa_0^2}{12 k_B T} \ . \tag{8}$$

The average time corresponding to a single step is $\langle \tau \rangle \approx 1/k(T)$ (the function $k(T)$ was described in Sec. IV). Based on equations (7) and (8), we obtain the mobility of the flake with the fixed commensurate orientation to be $\mu_c \approx a_0^2 k(T)/(12 k_B T) = D_c/(k_B T)$. The contribution of the considered drift mechanism to the total mobility of the free flake can, correspondingly, be found as $\mu_{c1} = \alpha_c \mu_c = \phi \mu_c / (1+\phi)$.

Let us now consider drift of the free flake through rotation of the flake to the incommensurate states. In the case of ballistic rotation ($\Delta\varphi/\omega_T \ll \tau_c'$), the translational motion of the flake before it returns to the commensurate states is accelerated. Thus, the average distance passed in a single step is given by $\langle l_F \rangle = 0.5 F \langle \tau_{rot}^2 \rangle / M$. Using the Maxwell-Boltzmann distribution for the angular velocity $\omega$ of the flake, we find (see S5, Ref. 62) that

$$\langle l_F \rangle \approx \frac{\Delta\varphi^2}{\omega_T^2} \frac{F}{M} ln\left(\frac{\omega_T \tau'_c}{\Delta\varphi}\right) \qquad (9)$$

On the basis of the values for $\langle l_F \rangle$ and $\langle \tau \rangle = \langle \tau_{st} \rangle + \langle \tau_{rot} \rangle$, the contribution of the proposed drift mechanism through rotation of the flake to the incommensurate states in the case of ballistic rotation to the total mobility of the free flake is seen to be

$$\mu_{ib} \approx \frac{\Delta\varphi}{\sqrt{\pi}\omega_T M (1+\phi(T))} ln\left(\frac{\omega_T \tau'_c}{\Delta\varphi}\right) = \frac{D_{ib}}{k_B T}. \qquad (10)$$

In the case of diffusive rotation ($\Delta\varphi/\omega_T \gg \tau'_c \approx \tau_c$), the average distance passed by the flake in a single drift step can be estimated as $\langle l_F \rangle = \mu_t F \langle \tau_{rot} \rangle$, where $\mu_t = D_t / k_B T = \tau_c / M$ is the maximum mobility of the flake on the graphite surface determined by friction. Taking into account that $\langle \tau_{st} \rangle / \langle \tau_{rot} \rangle = \phi(T)$, we find $\mu_{id} \approx \tau_c M^{-1} (1+\phi(T))^{-1} = D_{id}/(k_B T)$.

It is seen that the Einstein relation between the mobility and the diffusion coefficient of the flake is satisfied for all diffusion and drift mechanisms of the flake in the whole temperature range. Similar to the diffusion coefficient (see FIG. 6), the contribution of the proposed mechanism through rotation of the flake to the incommensurate states to the total mobility of the free flake is the most prominent at $T \sim (1 \div 3) T_{com}$.

## VI. CONTROL OF DIFFUSION AND DRIFT IN NEMS

As discussed above, rotation of the flake to the incommensurate states can lead to an increase of the diffusion coefficient and mobility of the flake by orders of magnitude. Therefore, it can be possible to control the diffusion coefficient and mobility of the flake with the help of an external force which affects the orientation of the flake. Let us consider the ways to control the diffusion coefficient and mobility of the flake using an external field and estimate the required field strength.

Recently it was proposed that relative motion of graphene layers can be used in graphene-based NEMS[2]. Diffusion and drift of NEMS components can be undesirable in the course of NEMS operation[41,46]. In this case, it might be useful to fix the commensurate orientation of the flake by an external force. On the other hand, stochastic behavior of nanosystems can be employed in Brownian motors[44]. In particular, start and stop of a Brownian motor based on the motion of the graphene flake can be realized by switching between the commensurate and incommensurate states of the flake.

We suggest that the orientation of the flake can be fixed with an electric field if the flake is functionalized so that it has a dipole moment. Let us consider the case where the field is directed so that alignment of the dipole moment of the flake along the field brings the flake to an incommensurate state. There is always a commensurate state of the flake for which the angle between the dipole moment of the flake and the electric field is $\varphi_E \leq \pi/6$. Under the action of the external field, this state is clearly has the lowest energy among the commensurate states of the flake. Rotation of the flake from this commensurate state to the incommensurate state for which the dipole moment of the flake is directed along the field is energetically favorable as long as the energy gain from the alignment of the dipole moment along the field exceeds the energy difference between the incommensurate and commensurate states

$$dE_e \left(1 - cos\, \varphi_E \right) > \varepsilon_{in} N ,  \qquad (11)$$

where $d$ is the dipole moment and $E_e$ is the electric field strength. At $E_e << \varepsilon_{in} N / d$, the influence of the electric field on the orientation of the flake can be neglected.

We have applied the previously used LDA approach to calculate dipole moments of functionalized graphene flakes. Flakes with opposite zigzag edges terminated with hydrogen and fluorine atoms are considered. The flakes are 22 Å in width and have 9 hydrogen and fluorine atoms, correspondingly. For the flakes of 20 and 10 Å length along the armchair direction (consisting of 180 and 90 carbon atoms), we find the dipole moments to be $d \approx 5.6 \cdot 10^{-29}$ C·m and $2.8 \cdot 10^{-29}$ C·m, respectively. It is seen that the dipole moment is proportional to the length of the flake. The calculated values of the dipole moments are in agreement with the result obtained for carbon nanotubes[42,43,45]. Using formula (11) for the angle $\varphi_E = \pi/6$ corresponding to the maximum energy gain from alignment of the dipole moments of the flakes along the field, we estimate the electric field strength required to fix the orientation of the considered flakes to be $E_e > 7.5\varepsilon_{in} N / d \sim 1.4$ V/nm.

The orientation of the flake can be also fixed with a magnetic field if a magnetic cluster is deposited on top of the flake. Let us estimate the magnetic field strength required to fix the flake in an incommensurate state assuming that the cluster is semispherical. If the diameter of the cluster is about $\sim a_0 N^{1/2}$, the number of atoms in the cluster can be estimated as $N_c \sim \pi a_0^3 N^{3/2} \rho_m / (12 m_m)$, where $\rho_m$ is the metal density and $m_m$ is the metal atomic mass. For iron, the atomic magnetic moment can reach $\mu_m \sim 3\mu_B$ [70], where $\mu_B = 5.3 \cdot 10^{-5}$ eV/(T·atom). Assuming that deposition of the metal cluster on the graphene flake does not lead to a significant modification of the potential energy relief of the flake, the magnetic field strength required to fix the flake in the incommensurate state can be estimated on the basis of the condition

$$\mu_m B N_c (1 - \cos\varphi_B) > \varepsilon_{in} N , \qquad (12)$$

where $\varphi_B \leq \pi/6$ is the angle of rotation from the most favorable commensurate state to the incommensurate state for which the magnetic moment of the cluster is directed along the field. This condition is reduced to

$$B > \frac{12\varepsilon_{in} m_m}{\pi \mu_m \rho_m a_0^3 N^{1/2} (1 - \cos\varphi_B)} . \qquad (13)$$

For the flake consisting of $N \sim 200$ atoms, the estimate gives $B > 4$ T.

## VII. INFLUENCE OF DEFECTS ON GRAPHENE DIFFUSION

Since real systems often incorporate structural defects, the study of the influence of defects on the diffusion characteristics of a graphene flake on a graphite surface is also of interest. Let us discuss the potential energy reliefs for the flakes containing structural defects and analyze the effect of the defects on the diffusion coefficient of the flake on the basis of the MD simulations.

Three types of defects retaining the flat structure of the flake are considered (see FIG. 7): a vacancy, a Stone–Wales defect and a pair of separated 5-7 defects (a single 5-7 defect causes bending of the flake). The vacancy and Stone–Wales defects do not destroy the commensurability between the graphene flake and graphite surface (see FIG. 7a,b). For the flake containing the vacancy, the AB-stacking in which is the vacancy of the flake is located over a carbon atom of the underlying graphene layer becomes slightly more energetically favorable (by $0.061\varepsilon_{com} = 6.3 \cdot 10^{-3}$ meV/atom for the considered flake consisting of 178 atoms) than the AB-stacking in which the vacancy is located over the center of a hexagon. Furthermore, the energy difference between the AA-stacking and the minimum energy AB-stacking increases (by $0.27\varepsilon_{com} = 2.7 \cdot 10^{-2}$ meV/atom) compared to the flake without defects. In the case of the Stone-Wales defect, there is also a small energy difference for configurations corresponding to the AB-stacking (by $0.04\varepsilon_{com} = 4.1 \cdot 10^{-3}$ meV/atom). The energy difference between the AA-stacking and the minimum energy AB-stacking slightly decreases (by $0.82\varepsilon_{com} = 8.5 \cdot 10^{-2}$ meV/atom) compared to the flake without defects. It is seen that both the vacancy and Stone-Wales defects lead to negligible changes of the potential energy relief of the graphene flake on the graphite surface, in agreement with the result of paper[8]. Thus, the diffusion mechanisms considered above are adequate for graphene flakes incorporating these two types of structural defects. Correspondingly, such flakes can be used in NEMS based on manipulation of the diffusion and drift characteristics of graphene flakes.

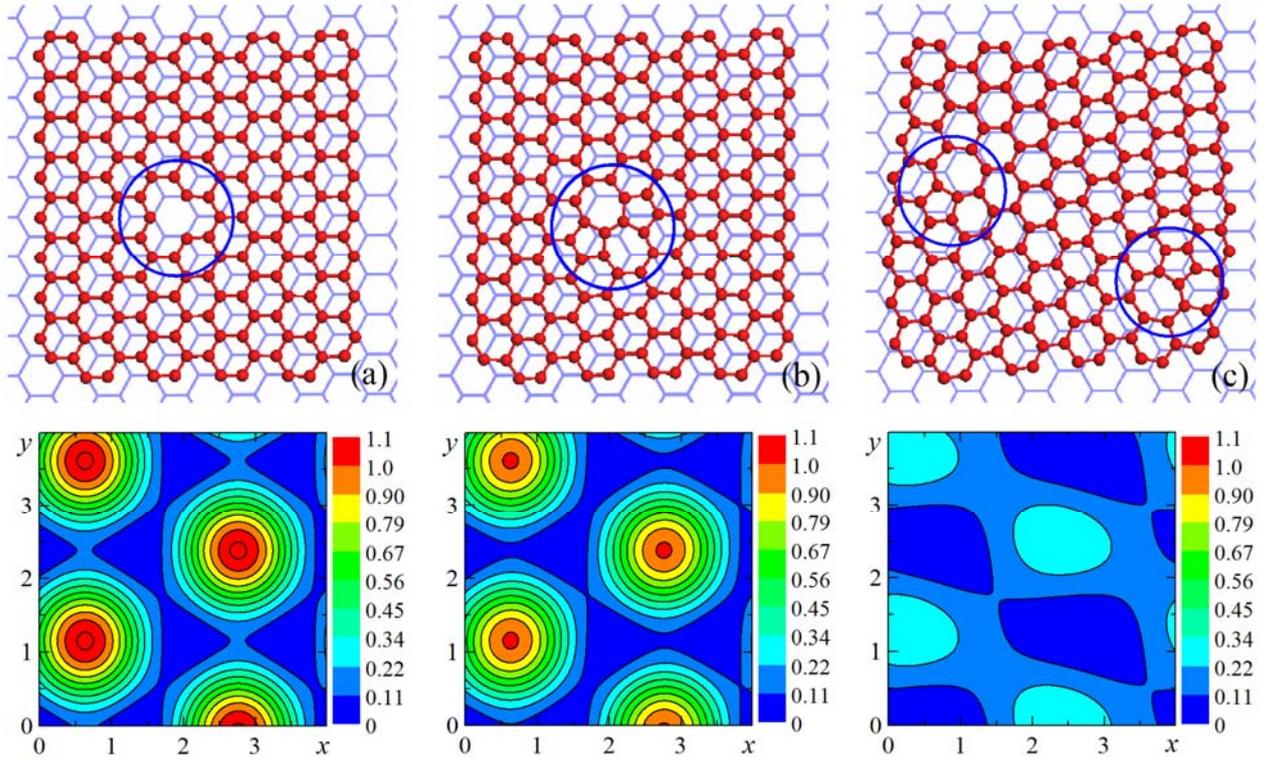

FIG. 7. Structures of graphene flakes with defects: (a) vacancy, (b) Stone-Wales defect, (c) pair of separated 5-7 defects. The interlayer interaction energies between the graphene flakes containing the defects and the graphene layer (in meV/atom) calculated using the Lennard–Jones potential as functions of the positions of the centers of mass of the flakes $x, y$ (in Å, $x$ and $y$ axes are chosen along the armchair and zigzag directions, respectively) for the commensurate states of the flakes ($\varphi = 0°$). The energies are given relative to the energy minima.

The pair of separated 5-7 defects leads to the loss of commensurability between the flake and the graphene layer and provides the potential energy relief which is much smoother than the potential energy reliefs for the flake without defects and for the flakes with the vacancy and Stone–Wales defects (see FIG. 7c). Furthermore, the magnitude of corrugation of the potential relief for this flake does not depend considerably on the flake orientation. So consideration of diffusion of the flake in terms of the commensurate and incommensurate states is not adequate for graphene flakes with 5-7 defects. It is clear that the diffusion coefficient of the flake incorporating the pair of separated 5-7 defects also weakly depends on the flake orientation. Therefore, control over the diffusion characteristics of such a flake by fixation of its orientation with the help of an external force is no longer possible.

TABLE III. Calculated diffusion coefficient $D$, average time $\langle \tau_{\text{rot}} \rangle$ of rotation by the angle $\Delta \varphi \approx \pi / 3$, average time $\langle \tau_{\text{st}} \rangle$ of stay in the commensurate states between these rotations and mean-square distance $\langle l^2 \rangle$ passed by the flake as it rotates by the angle $\Delta \varphi$ for the graphene flakes of different structure at temperature 300 K.

| Structure | $D$ ($10^{-4}$ cm$^2$/s) | $\langle \tau_{\text{st}} \rangle$ (ps) | $\langle \tau_{\text{rot}} \rangle$ (ps) | $\langle l^2 \rangle$ (Å$^2$) |
|---|---|---|---|---|
| without defects | $3.6 \pm 0.5$ | $13.4 \pm 1.2$ | $20.0 \pm 0.9$ | $240 \pm 30$ |
| vacancy | $2.5 \pm 0.9$ | $16.3 \pm 2.2$ | $19.0 \pm 1.2$ | $220 \pm 40$ |
| Stone-Wales defect | $2.7 \pm 0.9$ | $14.1 \pm 1.2$ | $21.9 \pm 1.1$ | $260 \pm 40$ |
| pair of separated 5-7 defects | $3 \pm 2$ | | | |

The smoother potential energy relief for the flake with defects may enhance its diffusion. However, defects can also increase the dynamic friction force between the flake and the underlying graphene layer (similarly to the effect found for double-walled carbon nanotubes[36-38]), which should reduce the diffusion coefficient. To reveal the actual effect of defects on the diffusion coefficient of the flake we have performed the MD simulations. The diffusion coefficient for each structure of the graphene flake is estimated on the basis of 4 MD simulations of 3.5-4 ns duration. The results of these calculations are presented in TABLE III. A considerably large spread in the data obtained for the flakes with the defects is related to an increase of the level of thermodynamic fluctuations in this case[36]. It is seen from TABLE III that the differences in the diffusion coefficients for all the considered flakes with defects and for the flake without defects are within the error bars of the calculations.

**VIII. CONCLUSION**

Different mechanisms of diffusion and drift of the graphene flake on the graphite surface are systematically analyzed in the wide ranges of temperatures and sizes of the flake. In addition to diffusion and drift of the flake by transitions between adjacent energy minima in the commensurate state, a new mechanism of anomalous fast diffusion and drift through rotation of the flake to the incommensurate states is proposed. The molecular dynamics simulations of diffusion of the free flake and of the flake with the fixed commensurate orientation are performed in the temperature range of 50–500 K. Analytic expressions for the diffusion coefficient and mobility of the flake are derived. Contributions of different diffusion and drift mechanisms to the diffusion coefficient and mobility of the flake are analyzed. Both the molecular dynamics simulations and estimates based on the analytic expressions demonstrate that the proposed diffusion mechanism is dominant at temperatures $T \sim (1 \div 3) T_{com}$. We believe that these results can be also applied to polycyclic aromatic molecules on graphene and should be qualitatively valid for a set of commensurate adsorbate-adsorbent systems.

We also suggest that the diffusion coefficient and mobility of the flake can be controlled via fixation of the flake orientation. This can be implemented with the help of an electric field if the flake possesses a dipole moment as a result of its functionalization or with the help of a magnetic field if a magnetic cluster is placed on the top of the flake. The proposed methods of control over diffusion and drift of graphene flakes can be used for elaboration of graphene-based NEMS.

The MD simulations show that the structural defects (Stone–Wales defect, vacancy and pair of separated 5-7 defects), which are often present in real systems, have the weak influence on the diffusion coefficient of the flake. However, the diffusion mechanisms are different for the flakes containing different type of defects. The dynamic behavior of the flake containing the vacancy or Stone–Wales defect is seen to be analogous to the behavior of the perfect flake. Namely, for these defects, the diffusion mechanism through rotation of the flake to the incommensurate states is dominant under the same conditions as for the flake without defects. Therefore, such flakes can be used in NEMS based on manipulation of the diffusion and drift characteristics of graphene flakes. For the flake containing the pair of 5-7 defects, the diffusion coefficient is found to weakly depend on its orientation. Therefore, for such flakes, control over the diffusion characteristics via fixation of the flake orientation is not possible.

From the analytic expressions derived here, it is seen that the diffusion coefficient and mobility of the graphene flake on the graphite surface depend exponentially on the difference in the interlayer energies of the commensurate and incommensurate states of the flake. Both *ab initio* and empirical calculations are shown to provide similar potential reliefs of the interlayer interaction energy for graphene (see Eq.(1)), which can be characterized with a single energy parameter. Therefore, we suggest that measurements of the temperature dependence for the diffusion coefficient of the graphene flake can also give a *true* value of the barrier for relative motion of graphene layers. For example, real-time TEM visualization[71] was recently applied to see single carbon and hydrogen adatoms as well as migration of carbon chains on graphene and dynamics of defects. Therefore, this technique can be used to measure diffusion coefficients of graphene flakes or polycyclic aromatic hydrocarbons on a graphene layer or a graphite surface. The knowledge of the barrier for relative motion of graphene layers is in particular valuable for interpretation of the data obtained using the friction force microscope[3–12] and for explanation of the dependence of thermal conductivity of few-layer graphene on the number of layers[72].

**ACKNOWLEGDEMENT**

This work has been partially supported by the RFBR grants 11-02-00604 and 10-02-90021-Bel.